\documentclass{emulateapj}

\usepackage{color}

\newcommand{\gr}{$\gamma$-ray}

\shorttitle{Detection of \gr{} emission from IC~310 by the MAGIC telescopes}
\shortauthors{Aleksi\'c et al.}

\begin{document}


\title{Detection of very high energy $\gamma$-ray emission from the Perseus cluster head-tail galaxy IC~310 by the MAGIC telescopes}


%
\author{
J.~Aleksi\'c\altaffilmark{a},
L.~A.~Antonelli\altaffilmark{b},
P.~Antoranz\altaffilmark{c},
M.~Backes\altaffilmark{d},
J.~A.~Barrio\altaffilmark{e},
D.~Bastieri\altaffilmark{f},
J.~Becerra Gonz\'alez\altaffilmark{g,}\altaffilmark{h},
W.~Bednarek\altaffilmark{i},
A.~Berdyugin\altaffilmark{j},
K.~Berger\altaffilmark{g}
% ...
\\
\vspace{0.1cm}
(\emph{The MAGIC Collaboration})\\
\vspace{0.2cm}
A.~Neronov \altaffilmark{y},
C. Pfrommer \altaffilmark{z},
A.~Pinzke\altaffilmark{1},
D.~V.~Semikoz \altaffilmark{2}
}
\altaffiltext{a} {IFAE, Edifici Cn., Campus UAB, E-08193 Bellaterra, Spain}
\altaffiltext{b} {INAF National Institute for Astrophysics, I-00136 Rome, Italy}
\altaffiltext{c} {Universit\`a  di Siena, and INFN Pisa, I-53100 Siena, Italy}
\altaffiltext{d} {Technische Universit\"at Dortmund, D-44221 Dortmund, Germany}
\altaffiltext{e} {Universidad Complutense, E-28040 Madrid, Spain}
\altaffiltext{f} {Universit\`a di Padova and INFN, I-35131 Padova, Italy}
\altaffiltext{g} {Inst. de Astrof\'{\i}sica de Canarias, E-38200 La Laguna, Tenerife, Spain}
\altaffiltext{h} {Depto. de Astrof\'{\i}sica, Universidad de La Laguna, E-38206 La Laguna, Spain}
\altaffiltext{i} {University of \L\'od\'z, PL-90236 Lodz, Poland}
\altaffiltext{j} {Tuorla Observatory, University of Turku, FI-21500 Piikki\"o, Finland}
% ...
\altaffiltext{y} {SDC Data Center for Astrophysics, Geneva Observatory, Chemin d'Ecogia 16, 1290 Versoix, Switzerland} 
\altaffiltext{z} {HITS, Schloss-Wolfsbrunnenweg 33, 69118 Heidelberg, Germany; CITA, University of Toronto, M5S 3H8 Toronto, ON, Canada}
\altaffiltext{1}  {Stockholm University, SE - 106 91 Stockholm, Sweden}
\altaffiltext{2}  {APC, 10 rue Alice Domon et Leonie Duquet, F-75205 Paris Cedex 13, France}
% ...




\begin{abstract}
We report on the detection with the MAGIC telescopes of very high energy \gr{}s from IC~310, a head-tail radio galaxy in the Perseus galaxy cluster, observed during the interval November 2008 to February 2010.
The \textit{Fermi} satellite has also  detected this galaxy. 
The source is detected by MAGIC at a high statistical significance of $7.6~\sigma$ in 20.6~hr of stereo data.
The observed spectral energy distribution is flat with a differential spectral index of $-2.00\pm0.14$. 
The mean flux above $300$~GeV, between October 2009 and February 2010, $(3.1\pm0.5)\times10^{-12}~\mathrm{cm^{-2}~s^{-1}}$, corresponds to $(2.5\pm0.4)$\% of Crab Nebula units.
Only an upper limit, of $1.9$\% of Crab Nebula units above $300$~GeV, was obtained with the 2008 data. 
This, together with strong hints ($>3\sigma$) of flares in the middle of October and November 2009, implies that the emission is variable.
The MAGIC results favour a scenario with the very high energy emission originating from the inner jet close to the central engine.  
More complicated models than a simple one-zone SSC scenario, e.g.~multi-zone SSC, external Compton or hadronic, may be required to explain the very flat spectrum and its extension over more than three orders of magnitude in energy.

\end{abstract}

\keywords{gamma rays: galaxies --- galaxies: active --- galaxies: individual (IC~310)}


\section{Introduction}
Most of the presently known extragalactic very high energy (VHE, above $300$~GeV) \gr{} emitters ($\sim$40) are blazars. 
So far only two radio galaxies, M~87 \citep{ah03} and Cen~A \citep{ah09}, and two starburst galaxies, NGC 253 \citep{ac09} and M82 \citep{k09}, have been clearly identified in this energy range. 

IC~310 (redshift z=0.019) is a head-tail radio galaxy located in the Perseus cluster at $0.6^\circ$ (corresponding to $\sim$ 1Mpc) from the cluster's central galaxy, NGC~1275.
Head-tail radio galaxies display a radio morphology consisting of a bright head, close to the optical galaxy and a fainter, elongated tail.
In the standard explanation, the jets are bent towards one direction creating the ``head" structure.
At larger distances they fan out in a characteristic tail that extends over many tens to hundreds of kpc. 
When the flow of intracluster medium (ICM) impacting these galaxies (in their rest frame) is supersonic (Mach number larger than $1$), the ram pressure of the ICM causes the jets to bend \citep{brb79}.
If the flow is transsonic (Mach number $\sim 1$), the thermal pressure gradient of the interstellar medium of these galaxies, due to their motion through the ICM, determines the bending \citep{jo79}.
In this last model, the inflow is decelerated and heated by a bow shock in front of the galaxy, which also generates a turbulent wake that re-accelerates the relativistic particle population in the tail and illuminates the tail.

The radio contours of IC~310 show an extended emission, pointing away from the direction of NGC~1275.
The length of this tail measured in radio varies between $0.12^\circ$ and $0.27^\circ$ \citep{sb98,lr05}.
The X-ray image of IC~310 observed by XMM-Newton is compatible with a point-like emission from the core and with no X-ray emission from its extended radio structure \citep{s05}.
Interestingly, \citet{s05} also showed that the X-ray emission may originate from the central AGN of the BL Lac-type object.
Other observed characteristics of IC~310 (e.g.~no strong emission lines, spectral indexes in radio and X-ray) suggest that it may also be a dim (weakly beamed) blazar \citep{rsp99}.

The LAT instrument on board the {\it Fermi} satellite \citep{fermi_lat} has recently detected IC~310 \citep{nsv10} with 5 (3) photons above $30$~GeV  (above $100$~GeV).
At lower energy (i.e.~from $100$~MeV to $1$~GeV), only the cluster's central galaxy, NGC~1275, is visible \citep{fermi_ngc, nsv10}.

In this letter, we present the results of recent observations of the Perseus cluster performed with the Major Atmospheric Gamma Imaging Cherenkov (MAGIC) telescopes and the detection of IC~310 (ATel~\#2510).

\section{Observation and Analysis}
MAGIC consists of two 17~m Imaging Atmospheric Cherenkov Telescopes (IACT) located at the Roque de los Muchachos, Canary Island of La Palma ($28^\circ$N, $18^\circ$W), at the height of 2200 m a.s.l.
The commissioning of the second MAGIC telescope was finished in 2009, and since the end of 2009 both telescopes work together in stereo mode \citep{cortina09}.
The stereo observations provide an excellent sensitivity of $<1$\% of the Crab flux (C.U.)\footnote[1]{Hereafter C.U. stands for Crab Units, defined as the fraction of the Crab Nebula flux defined by Eq.~2 of \citet{al08}, that corresponds, e.g.~for energies above $300$~GeV, to $12.4\times10^{-11}$~cm$^{-2}$~s$^{-1}$.} in the medium energy range in 50~hr of observations \citep{co09}.

The MAGIC~I telescope was used to observe the Perseus cluster for a total of 94~hr between November 2008 and February 2010.
The analysis of 33.4~hr of data taken in 2008 was presented in \citet{al10}; it focused on the physics of the Perseus cluster and NGC~1275.
The skymap presented in \citet{al10} does not show significant excesses in the IC~310 position\footnote[2]{Throughout this paper, we refer to the IC~310 position as the position reported by the NASA Extragalactic Database (NED) from \cite{2mass03} catalog}.
Since the end of October 2009 the second MAGIC telescope was also taking data, allowing the stereoscopic analysis.

Observations were performed in the so-called wobble mode \citep{fomin}, with data equally split in two pointing positions offset by $0.4^\circ$
from the direction of NGC~1275. 
IC~310  was in the field of view at the angular distance of  $0.25^\circ$ and $1^\circ$ for individual wobble positions. 
Since the \gr{} collection area in the latter case is significantly lower (by a factor of $\sim3$), only data with $0.25^\circ$ offset were used for the signal search and for obtaining the spectrum and the light curve. 

We separately analyzed the MAGIC~I single telescope (hereafter mono) and stereo data. 
The mono and stereo are only partially independent systems, differing in the analysis method, thus there can be some residual systematic error between them. 
For single telescope observations, the systematic error in the determination of the flux is $\sim 30\%$ \citep{al08}.
Additionally, IC~310 was not observed in the standard wobble observation mode; this increases the systematic error from the background estimate. 
The mono data are treated with the standard MAGIC~I analysis chain \citep{al08,aliu09}.
In the analysis of the stereo data we took advantage of the impact parameters with respect to each telescope and the height of the shower maximum.
Those stereo parameters improve the gamma/hadron separation and the energy reconstruction (\citealp{ah97}; MAGIC stereo performance paper, in prep.). 
We used a new method to reconstruct  the arrival directions that improves the angular resolution and the sensitivity of the MAGIC telescopes. 
The method is based on the ``DISP RF" technique (see \citealp{al10b} for details), adapted to the stereo observations.
The estimate of an arrival direction is performed independently for each telescope using the shape and time information of a corresponding image. 
In order to obtain the best performance, those are combined with the crossing point of the main axes of the images from both telescopes.

\begin{figure}[t]
\centering 
\includegraphics[width=0.4\textwidth]{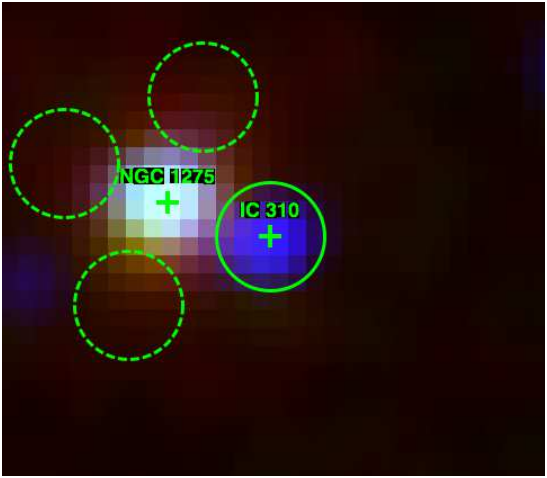}
\caption{
\textit{Fermi}-LAT count rate image of the IC~310 region in three energy bands: $0.3-3$~GeV (red), $3-30$~GeV (green) and $30-300$~GeV (blue).
The position of IC~310 and NGC~1275 are shown with green crosses. 
Solid (dashed) circles show the signal (background) integration regions used for obtaining the \textit{Fermi}-LAT spectrum.
} \label{fermi_counts}
\end{figure}

We also analyzed \textit{Fermi} data taken during the period between 2008 August 4 and 2010 July 15 following the approach of \citet{nsv10}.
The data were filtered with {\it gtselect} tool and the \textit{Fermi} exposure at the source position was calculated using {\it gtexposure} tool\footnote[3]{\label{footnote3}See http://fermi.gsfc.nasa.gov/ssc/data/analysis/scitools/ for details on the \textit{Fermi} analysis, we used the software version v9r15p2.}.
Only events of ``diffuse'' class were retained in the analysis. 
We obtained the spectrum of IC~310 in the $2-200$~GeV energy band in two different ways. 
First, we performed a spectral fit using the standard \textit{Fermi} unbinned likelihood analysis, taking into account all sources from the $1^{st}$ year \textit{Fermi} catalog \citep{fermi_cat} within $10^\circ$ of IC~310. 
We then obtain spectral points by extracting photon counts from the circle of radius $0.3^\circ$ centered on the source. 
Taking into account the proximity to a bright, nearby source (radio galaxy NGC~1275), we estimated the background by taking three apertures at $0.6^\circ$ away from NGC~1275 (see Fig.\ref{fermi_counts}).

\section{Results}
\begin{figure*}[t]

\includegraphics[width=0.32\textwidth]{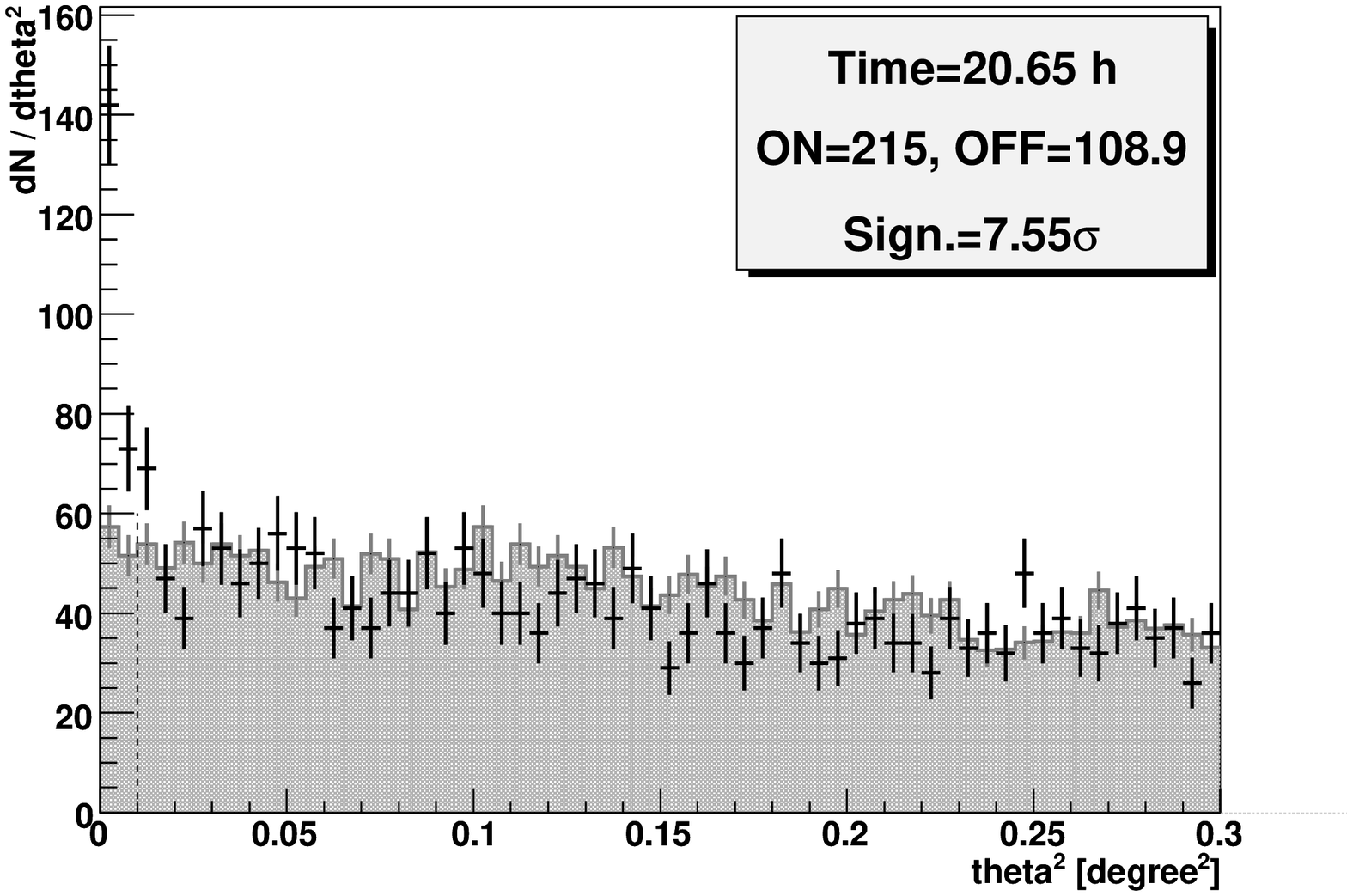}\hspace{-4.7cm}\makebox[4.7cm][l]{\raisebox{3.4cm}{\sf \scriptsize STEREO 2009-2010}}
\includegraphics[width=0.32\textwidth]{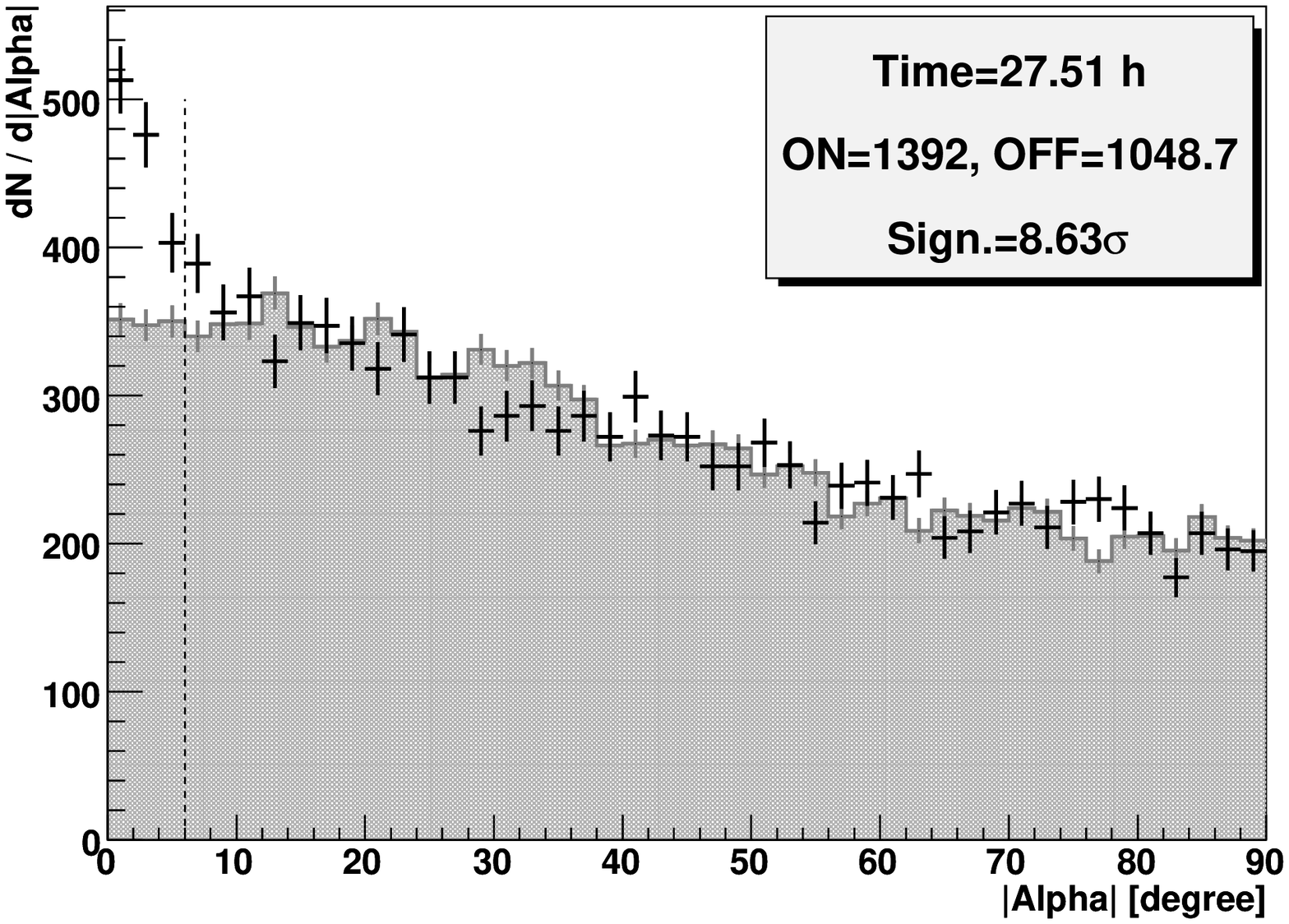}\hspace{-4.5cm}\makebox[4.5cm][l]{\raisebox{3.4cm}{\sf \scriptsize MONO 2009-2010}}
\includegraphics[width=0.32\textwidth]{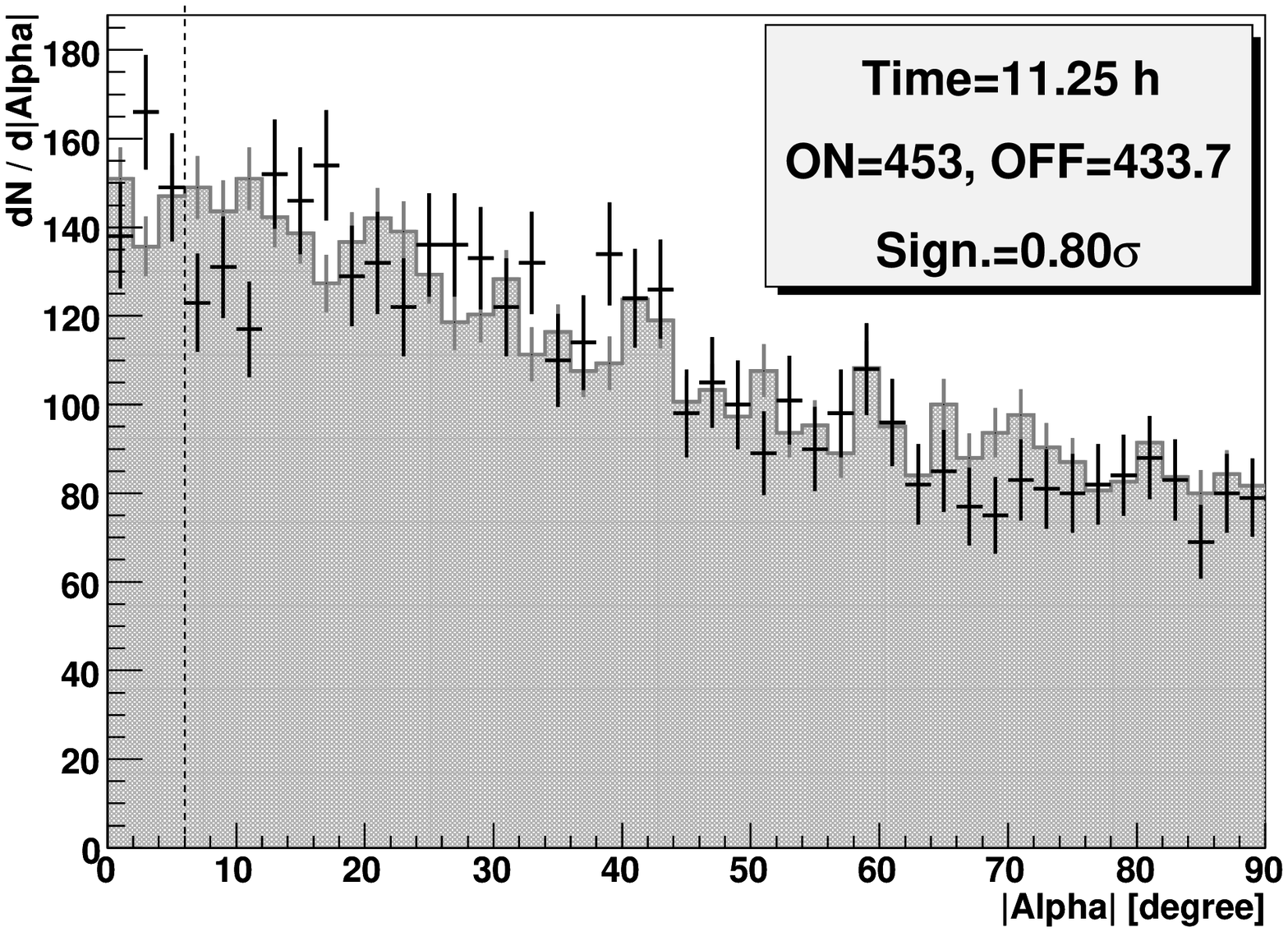}\hspace{-4.3cm}\makebox[4.3cm][l]{\raisebox{3.4cm}{\sf \scriptsize MONO 2008}}
\caption{
Theta$^2$ distribution of the IC~310 signal and background estimate from stereo observations taken between October 2009 and February 2010 (left panel).
Alpha distribution from mono observations taken between September 2009 and February 2010 (middle panel), and November-December 2008 (right panel).
Only the pointing position $0.25^\circ$ away from IC~310 is used.
The cuts result in an energy threshold (defined as a peak of the differential MC energy distribution) of $\sim260$~GeV for both mono and stereo data (see the text for details).
} \label{theta2plot}
\end{figure*}
After the data quality check, we obtained a sample for the period from October 2009 to February 2010, of 20.6~hr of MAGIC stereo data. 
The theta$^2$ (squared distance between true and reconstructed source position; see e.g.~\citealp{daum97}) distribution of the signal coming from IC~310 and the background estimation are shown in Fig.~\ref{theta2plot} (left panel). 
We found an excess of 106 events, corresponding to a $7.6$~$\sigma$ significance (calculated according to the prescription by \citealp{lm83}, Eq.17).

The source was also detected in the 27.5~hr of mono data (September 2009 -- February 2010) with a significance of $8.6$~$\sigma$. 
Note that since part of the MAGIC~I data set are also used in the stereo analysis, the two significances are not completely independent.
The corresponding alpha (the angle between the main axis of the \gr{} induced shower image and direction to the true source position; see e.g.~\citealp{al08}) distribution is also shown Fig.~\ref{theta2plot} (middle panel).
The  different signal significance obtained in stereo ($7.6\sigma$) is similar to the one of mono scaled to the same observation time ($8.6\sigma \times\sqrt{20.6~\mathrm{hr} / 27.5~\mathrm{hr}}=7.5\sigma$).  
This is because the mono data have been taken over a longer time period, including a higher emission state in October 2009 (see below). 
Moreover, in significance calculated according to \citet{lm83} the background is estimated using both ON and OFF measurements.
Thus for a high signal/background ratio (as in the case of excellent gamma/hadron separation obtained in stereo observations), the background is overestimated and this lowers the significance. 
On the other hand a simple calculation of $\mathrm{excess/\sqrt{OFF}}$ scaled to the same observation time gives a higher value for stereo, $10.2$, than for mono, $9.2$, observations.

It is interesting to note that the 11.2~hr of good quality, mono data taken at the end of 2008 do not show any significant excess at the position of IC~310 (see Fig.~\ref{theta2plot}, right panel).
These data yield an upper limit for the flux $F(>300\mathrm{GeV})<1.9\%$~C.U. (calculated using the \citealp{rolke} method with 95\% of confidence level and assuming 30\% systematic error in absolute flux level).

\begin{figure}[t]
\centering 
\includegraphics[width=0.5\textwidth]{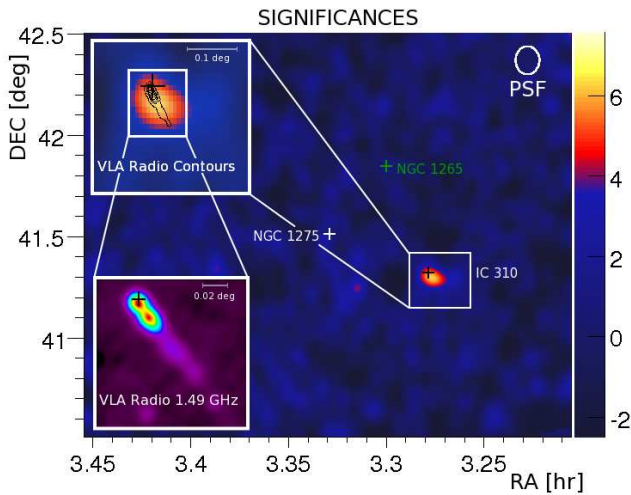}
\caption{
Significance skymap from the MAGIC stereo observation (42~hr; both pointing positions) for energies above $400$~GeV. 
We also show an enlargement of the IC~310 region overlaid with the NVSS (NRAO VLA Sky Survey at 1.49~GHz; \citealp{nvss}) contours (top left inserted panel) and the corresponding NVSS image (bottom left inserted panel).
The NVSS data were obtained with \emph{Aladin} \citep{aladin}.
Positions of IC~310, NGC~1275  and NGC~1265 are marked with black, white, and green crosses respectively.
} \label{skymap}
\end{figure}

At energies above $400$~GeV the MAGIC telescopes working together in stereo have a point spread function (PSF) of $\sim0.06^\circ$, defined as a 40\% containment radius, corresponding to a $\sigma$ of a 2-dimensional Gaussian. 
In Fig.~\ref{skymap}, we show the significance map of the Perseus cluster region above $400$~GeV. 
The bright  spot is consistent with the position of the IC~310. 
In the panels inserted in Fig.~\ref{skymap}, we also show archival (non-simultaneous) IC~310 VLA radio data \citep{nvss}.

\begin{figure}[t]
\centering
\includegraphics[width=0.45\textwidth]{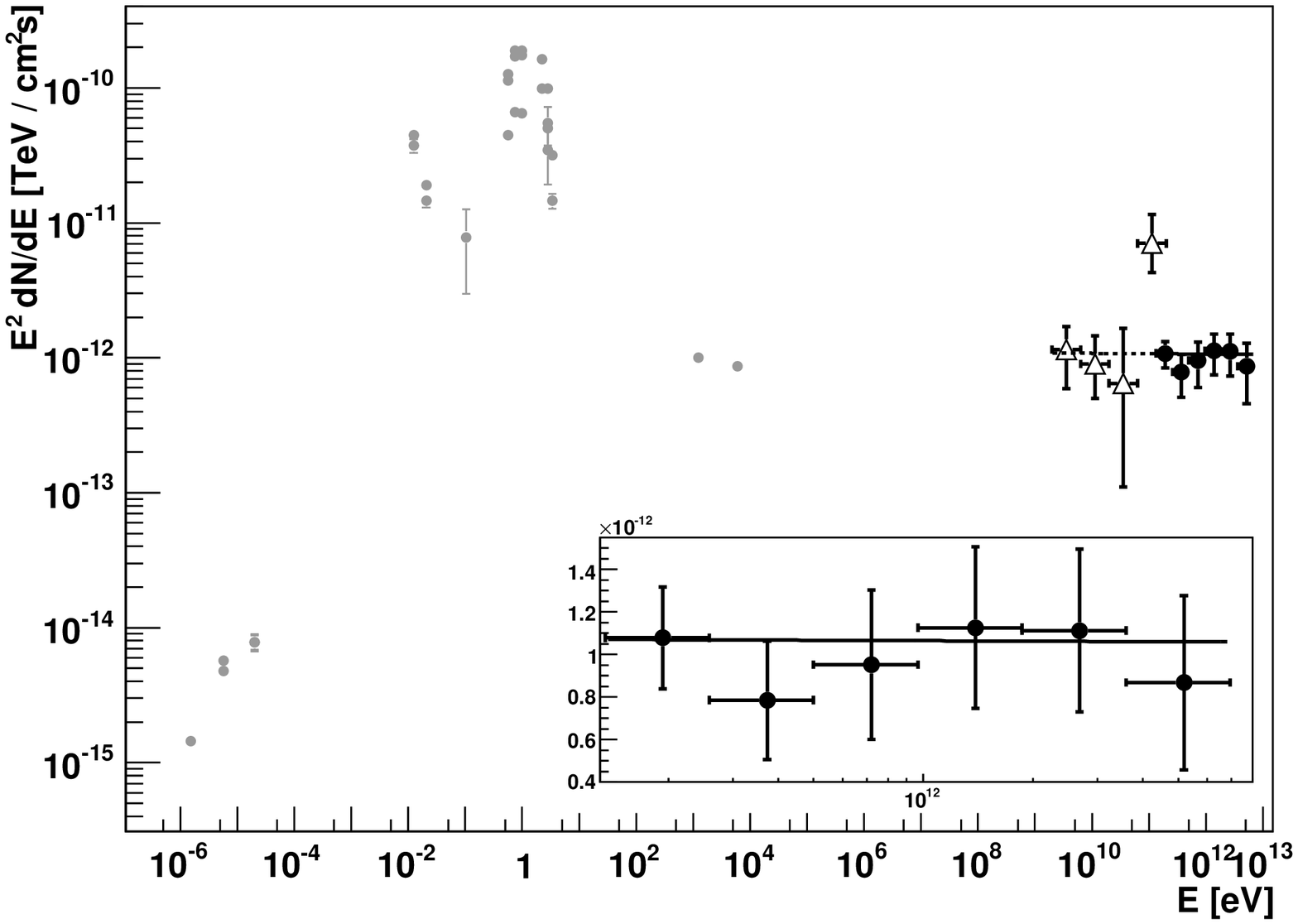}
\caption{
SED of IC~310 obtained with 20.6~hr of the MAGIC stereo data (full circles). 
Open triangles show the flux measurements from the \textit{Fermi}-LAT from its first two years of operation. 
Archival X-ray \citep{s05}, 
optical \citep{kw68, v91}, 
IR \citep{2mass03,k89, iras88}
and radio \citep{gc91, b91, ccb02, wb92, d96} data
obtained from the NED database are shown with grey dots. 
The solid line shows a power law fit to the MAGIC data, and the dotted line is its extrapolation to GeV energies.
We also show a zoom-in of the MAGIC points.
} \label{spectrum}
\end{figure}

The MAGIC stereo observations reveal a flat spectral energy distribution (SED)  between 150~GeV and 7~TeV without any visible curvature or cut-off (see Fig.~\ref{spectrum}). 
The differential flux in units of $\mathrm{cm^{-2}s^{-1}TeV^{-1}}$ is well described ($\chi^2/n_{dof}=2.3/4$) by a pure power law:
\begin{equation}
dN/dE = (1.1\pm0.2) \times 10^{-12} (E/\mathrm{TeV})^{-2.00\pm0.14}.
\end{equation}
The mean \gr{} flux above $300$~GeV obtained from the stereo observations between October 2009 and February 2010 is $(3.1\pm0.5)\times10^{-12}~\mathrm{cm^{-2}~s^{-1}}$, corresponding to $(2.5\pm0.4)\%$~C.U.
Comparing this with the upper limit from the 2008 data suggests variability of IC~310 on a one-year time scale.

\begin{figure}[t]
\centering 
\includegraphics[width=0.45\textwidth]{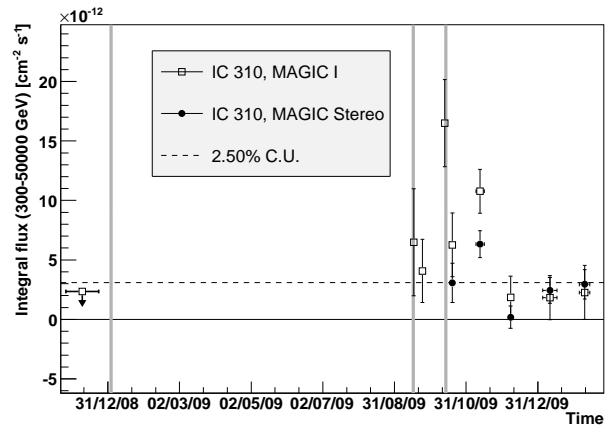}
\caption{
Light curve (in 10-day bins) of the \gr{} emission above $300$~GeV obtained with the mono (open squares) and the stereo (full circles) MAGIC data. 
The open square with an arrow is the upper limit on the emission in November-December 2008. 
Vertical grey lines show the arrival times of $>100$~GeV photons from the \textit{Fermi}-LAT instrument.
The horizontal dashed line is a flux level of 2.5\% C.U. 
}\label{lightcurve}
\end{figure}

The light curves of IC~310 \gr{} emission above 300~GeV obtained both with the mono and stereo data are presented in Fig.~\ref{lightcurve}.
%
Hints of variability can be seen in the data.
Fitting the individual light curves assuming constant flux yields $\chi^2 / n_{dof}$ = $27.6 / 7$ (for mono, corresponding to $3.5\sigma$ ) and $17.5 / 4$ (for stereo, corresponding to $3.0\sigma$). 
The largest deviations from the mean value are for  the intervals $13-14$ October 2009 ($3.1$~$\sigma$ in mono), and $9-16$ November 2009 ($3$~$\sigma$ in mono, $3.2$~$\sigma$ in stereo).

Until February 2010, the \textit{Fermi}-LAT instrument observed only three photons with energies above $100$~GeV from the direction of IC~310 \citep{nsv10}.
It is interesting to note that one of those \gr{}s was observed on 15th of October, nearly coincident with the higher flux seen in mono. 

The standard \textit{Fermi} likelihood analysis gave a ``Test Statistics'' value of 79 from IC~310 above 1~GeV (corresponding to a $\sim 9$~$\sigma$ detection)
\footnote[4]{Note that this significance is not calculated according to the \citet{lm83} method, see \textit{Fermi} analysis description (footnote~\ref{footnote3}).}. 
Assuming a simple power law for the spectrum, we obtain the differential flux 
$dN/dE = (9.5\pm2.9) \times 10^{-9}\, (E/\mathrm{10 GeV})^{-1.58\pm0.25} \mathrm{cm^{-2}\,s^{-1}\,TeV^{-1}}$.
The \textit{Fermi} spectral index is very hard, mostly due to the last point.

\section{Conclusions and Discussion}

The MAGIC telescopes have detected VHE \gr{} emission with high statistical significance from the direction of IC~310 with a mean \gr{} flux above $300$~GeV, between October 2009 and February 2010, of $(2.5\pm0.4)\%$~C.U.
The source seems to be variable on the time scales from weeks to about one year. 
No significant excesses were observed in the 2008 mono data and flares are possibly present at the $\gtrsim3$~$\sigma$ level in October and November 2009. 
The stereo observations yield a spectrum that is well fitted by a simple power law with a spectral index of $-2.00\pm0.14$.

The MAGIC angular resolution is not sufficient to determine the location of the VHE emission region within the radio galaxy.
Therefore, it is not clear whether the observed \gr{} emission is connected with the tail of the source or if it is produced at the base of the jet, close to the central engine of the source (as in blazars). 
The strong indications of variability disfavor the \gr{} production at the bow shock, discussed by \citet{nsv10}, because in this case the emission should be steady on time scales of thousands of years. 
Variability with a time scale of a year (a week) constrains the size of emission region to be $\lesssim 10^{18}$~cm  ($\lesssim 2\times 10^{16}$~cm) across, assuming no Doppler boosting of the flux, which is much smaller than the total size of the tail  ($\sim10^{24}$~cm).
Additionally, one can estimate the mass of the central black hole, $M_{\rm BH}$,  of an active  galaxy using the correlation between black hole mass and the central velocity dispersion of the host galaxy \citep{tr02}. 
The measured velocity dispersion in IC~310 ($230$~km/s, \citealp{me95}) yields $M_{BH}=2.4\times 10^8 M_\odot$, corresponding to a Schwarzschild radius of $R_{SH}=7\times 10^{13}$~cm.  
This indicates that the most probable location of the \gr{} emission region is in the innermost part of the jet \citep[as e.g.~for M~87, see][]{acc09}.

The extrapolation of the IC~310 spectrum obtained with the MAGIC telescopes is in good agreement with the \textit{Fermi} spectrum below 60~GeV.
On the other hand, there is a large deviation in the last energy bin measured by \textit{Fermi} (see Fig.~\ref{spectrum}).
The \gr{} flux from our observations in this energy bin predicts 0.6 photons, while 4 photons were observed by \textit{Fermi}-LAT.
Assuming a Poisson distribution, the probability of obtaining $\ge 4$ photons is $3.4\times 10^{-3}$ (corresponding to $2.7$ standard deviations) 
so the discrepancy may be a statistical fluctuation.
However, the \textit{Fermi}-LAT, having an energy resolution of $\sim10\%$, observed 3 of the 4 photons with nearly the same energies (98.5, 105, and 111~GeV).
If confirmed by future observations, these events may indicate the presence of a peculiar peak or bump in the IC~310 spectrum given that the remaining \textit{Fermi}-LAT data agree with the MAGIC measurement. 
The detection of such a relatively narrow feature in the spectrum of a radio galaxy may be a clue regarding the particle acceleration mechanism at the base of AGN jets \citep[e.g.~``direct'' gamma-ray 
emission during acceleration of particles, ][]{neronov07}.  This remains a suggestion since 
the source seems to be variable, and the MAGIC and \textit{Fermi} data used here were not taken simultaneously.

The combined MAGIC and \textit{Fermi} spectrum (besides the above mentioned bump) is consistent with a flat $E^{-2}$ spectrum stretching without a break over more than $3$ orders of magnitude in energy (2~GeV  -- 7~TeV). 
This is similar to the flat VHE spectra of M 87, another radio galaxy detected at TeV energies \citep{ah06,al08_M87,acc08}.
Such an extended, flat $E^{-2}$ spectrum is hard to obtain in a simple one-zone SSC model (\citealp{r67}; \citealp{mgc92}). 
Instead, a viable model of emission might be inverse Compton scattering of external IR photon background photons from accretion flow or from the inner jet (see e.g.~\citealp{neronov07}). 
Alternatively, a flat spectrum can be produced in the hadronic models (e.g.~the proton blazar model, \citealp{m93}).
In more complicated, multi-zone leptonic models, the GeV-TeV emission of a few slightly shifted inverse Compton peaks can also emulate a flat spectrum \citep[e.g.~spine-sheath layer model, ][]{tg08}. 

Finally, using the model by \citet{ebl},  we find that the change in the spectrum due to the absorption in the extragalactic background light radiation field is within the error of the spectral slope.

IC~310 and NGC~1265 were the first two head-tail galaxy discovered by astronomers \citep{rw68}. 
Now, IC~310 is the first head-tail radio galaxy detected in the VHE \gr{}s by ground based telescopes as well as by the \emph{Fermi} satellite.
Additionally, it is also the first source discovered above $300$~GeV by the MAGIC telescopes working together in stereo mode.
This detection is important also for the VHE study of the Perseus cluster of galaxies.  
It may imply a substantial injection of high energy particles in the ICM by a non-central AGN.
Those particles might have important consequences on the possible cluster \gr{} emission due to cosmic-ray acceleration \citep{al10}.
The VHE IC~310 detection significantly enriches our knowledge of the \gr{} universe.

\acknowledgments
\section*{Acknowledgments}
We would like to thank the Instituto de Astrof\'{\i}sica de
Canarias for the excellent working conditions at the
Observatorio del Roque de los Muchachos in La Palma.
The support of the German BMBF and MPG, the Italian INFN, 
the Swiss National Fund SNF, and the Spanish MICINN is 
gratefully acknowledged. This work was also supported by 
the Marie Curie program, by the CPAN CSD2007-00042 and MultiDark
CSD2009-00064 projects of the Spanish Consolider-Ingenio 2010
programme, by grant DO02-353 of the Bulgarian NSF, by grant 127740 of 
the Academy of Finland, by the YIP of the Helmholtz Gemeinschaft, 
by the DFG Cluster of Excellence ``Origin and Structure of the 
Universe'', and by the Polish MNiSzW Grant N N203 390834.
This research has made use of the NASA/IPAC Extragalactic Database (NED) which is operated by the Jet Propulsion Laboratory, California Institute of Technology, under contract with the National Aeronautics and Space Administration.


\end{document}